\documentstyle[epsf]{kluwer}
\begin{opening}
\title{Robustness of truncated $\alpha \Omega$ dynamos with a dynamic $\alpha$}
\author{Eurico \surname{Covas}\thanks{e-mail: E.O.Covas@qmw.ac.uk}}
\author{Andrew \surname{Tworkowski}\thanks{e-mail:A.S.Tworkowski@qmw.ac.uk}}
\author{Reza   \surname{Tavakol}\thanks{e-mail: reza@maths.qmw.ac.uk}}
\institute{Astronomy Unit, School of Mathematical Sciences,
Queen Mary and Westfield College, Mile End Road, London E1 4NS, UK.}
\author{Axel   \surname{Brandenburg}\thanks{e-mail:Axel.Brandenburg@ncl.ac.uk}}
\institute{Department of Mathematics and Statistics, University of Newcastle
upon Tyne, Newcastle NE1 7RU, UK.}
\end{opening}
\begin{document}
\begin{ao}
Web address: {\it http://www.maths.qmw.ac.uk/$\sim$eoc}
\end{ao}
%______________________________________________________________________
\begin{abstract}
%______________________________________________________________________
In a recent work (Covas et al. 1996),
the behaviour and the robustness of
truncated $\alpha\Omega$ dynamos with a dynamic
$\alpha$ were studied with respect to a number
of changes in the driving term of the dynamic $\alpha$ equation,
which was considered previously by Schmalz and Stix (1991) to
be of the form $\sim A_{\phi}B_{\phi}$.
Here we review and extend our previous work and consider the
effect of adding a quadratic quenching term of the form
$\alpha|{\bf B}|^2$. We find
that, as before, such a change can have significant effects
on the dynamics of the related truncated systems.
We also find intervals of (negative) dynamo numbers,
in the system considered by Schmalz and Stix (1991),
for which there is sensitivity with respect to
small changes in the dynamo number and the initial conditions,
similar to what was found in our previous work.
This latter behaviour may be of importance in producing
the intermittent type of behaviour observed in the sun.

%______________________________________________________________________

\end{abstract}

\keywords{Sun and stars; magnetic fields, mean field dynamos,
nonlinear dynamics, fragility, chaos}

%______________________________________________________________________
\section{Introduction}
%______________________________________________________________________
Given the importance of hydrodynamic dynamos in 
accounting for solar and stellar variabilities,
a large number of studies have been made of their 
dynamical modes of behaviour using a range 
of dynamo models. However,
the complexity and the numerical costs of employing the full
magneto-hydrodynamical partial differential equations
(Gilman 1983) have resulted in a great deal of effort going into the study 
of simpler related systems, including mean field
and truncated models.

Both the mean field models (cf. Krause and R\"adler 1980;
Brandenburg et al. 1989; Brandenburg, Moss and Tuominen 1989; 
Tavakol et al. 1995) and truncated systems 
(Zeldovich, Ruzmaikin and Sokoloff 1983;
Weiss, Cattaneo and Jones 1984; Feudel, Jansen and Kurths 1993) 
have been shown to be capable of producing 
a spectrum of dynamical modes of behaviour, including
equilibrium, periodic and chaotic states.

Here we shall concentrate on truncated models and recall that
there have been two approaches in the study of such 
models: the quantitative study of
the truncated equations resulting from concrete
partial differential equations (e.g. Schmalz and Stix 1991)
and the qualitative approach involving the use of
normal form theory to construct robust minimal low order 
systems which capture important aspects of the dynamo
dynamics (Tobias, Weiss and Kirk 1995).

Given the approximate nature of the truncated models,
an important question arises as to the extent to which the
results produced by such models are an artefact
of their details. 
This is potentially of importance since, on the basis of results
from dynamical systems theory, structurally stable systems
are not everywhere dense in the space of dynamical systems
(Smale 1966), and therefore small changes in the models can
potentially produce qualitatively important changes in their dynamics
(Tavakol and Ellis 1988; Coley and Tavakol 1992; Tavakol et al. 1995).

To partially answer this question,
we take a somewhat complementary approach to the 
previous works, by looking at the robustness
of the truncated models studied by Schmalz and Stix
with respect to a number of reasonable changes to their
components. This work extends and reviews our
previous work (Covas et al. 1996).
%______________________________________________________________________
\section{Mean field dynamo models with a dynamic $\alpha$}
%______________________________________________________________________
As an example of how the details of truncated models can
change their behaviour, we take the
truncated dynamic $\alpha$ models considered in Schmalz and Stix.
The starting point of this work is the
mean field induction equation
%______________________________________________________________________
\begin{equation}\label{induction}
\frac{\partial{\bf B}}{\partial t}=\nabla\,\times\,({\bf v}\,\times\,{\bf B}+
\alpha{\bf B}-\eta_t\nabla\,\times\,{\bf B}),
\end{equation}
%______________________________________________________________________
in the usual notation with the 
turbulent magnetic diffusitivity $\eta_t$ and 
the coefficient $\alpha$ arising 
from the correlation of small scale
(turbulent) velocity and magnetic fields (Krause and R\"adler 1980).

For the sake of comparison
we take an axisymmetrical configuration with one spatial dimension $x$ 
corresponding to a latitude coordinate
(and measured in terms of the stellar radius $R$), together with a
longitudinal velocity
with a constant radial gradient (the vertical shear $\omega_0$).
The magnetic field is given by
%______________________________________________________________________
\begin{equation}\label{Bspherical}
{\bf B}=\left(0,B_{\phi},\frac{1}{R}\frac{\partial A_{\phi}}{\partial x}\right),
\end{equation}
%______________________________________________________________________
where
$A_\phi$ is the $\phi$--component (latitudinal) of the magnetic vector
potential and $B_\phi$ the $\phi$--component of ${\bf B}$.
These assumptions allow equation (\ref{induction}) to be split into
%______________________________________________________________________
\begin{eqnarray}
\label{p1}
\frac{\partial A_{\phi}}{\partial t}&=&\frac{\eta_t}{R^2}
\frac{\partial^2A_{\phi}}{\partial x^2}+\alpha B_{\phi},\\
\label{p2}
\frac{\partial B_{\phi}}{\partial t}&=&\frac{\eta_t}{R^2}\frac{\partial^2 B_{\phi}}
{\partial x^2}+\frac{\omega_0}{R}\frac{\partial A_{\phi}}{\partial x}.
\end{eqnarray}
%______________________________________________________________________
Furthermore $\alpha$ is divided into a static (kinematic) and a dynamic
(magnetic) part as follows:
$\alpha=\alpha_0\cos x-\alpha_M(t)$, with its time-dependent part
$\alpha_M(t)$ satisfying an evolution equation of the form
%______________________________________________________________________
\begin{equation}\label{dynamicalpha}
\frac{\partial \alpha_M}{\partial t}= {\cal O}(\alpha_M) + {\cal F}({\bf B}),
\end{equation}
%______________________________________________________________________
where ${\cal O}(\alpha_M)$ is the damping term
and ${\cal F}({\bf B})$ the driving term, which
is a pseudo-scalar and quadratic in the magnetic field (see Covas et al. (1996) 
for details of this and the truncations of equations (\ref{p1}), (\ref{p2})
and (\ref{dynamicalpha})). We consider the interval $0\le x\le\pi$
(which corresponds to the full range of latitudes) and take $A=B=C=0$ 
at $x=0$ and $x=\pi$ as boundary conditions. We restrict the allowed modes
to the antisymmetric subset, that is, those modes that satisfy the condition ${\bf B}=0$
at $x=\pi/2$. 

Schmalz and Stix then fix the functional form of ${\cal F}$ and study  the
resulting modal truncations. Now given the fact that 
the exact form of ${\cal F}({\bf B})$ is not known precisely,
we shall examine in the next section 
how robust their results are with respect to changes  in
${\cal F}({\bf B})$.
%______________________________________________________________________
\section{Robustness with respect to changes in the driving term}
%______________________________________________________________________

In their studies, Schmalz and Stix
take the following form for the feedback term
%______________________________________________________________________
\begin{equation}
{\cal F}({\bf B})\sim A_{\phi}B_{\phi},
\label{ab}
\end{equation}
%______________________________________________________________________
and look at the various $N-$modal truncations of
these equations to find out what happens to the dynamical
behaviour of the resulting systems as $N$ is increased.
In all of the following discussion, we assume pure antisymmetric modes.

To determine the nature of the dynamics, we shall
employ the spectrum of Lyapunov exponents, and distinguish
signatures of the types $(-,-,-,\ldots)$, $(0,-,-,\ldots)$,
and $(+,0,-,\ldots)$ as corresponding to equilibrium, periodic,
and chaotic regimes respectively.

%______________________________________________________________________
\begin{figure}
%\vspace{7.2cm}
\centerline{\def\epsfsize#1#2{0.45#1}\epsffile[-50 60 800 580]{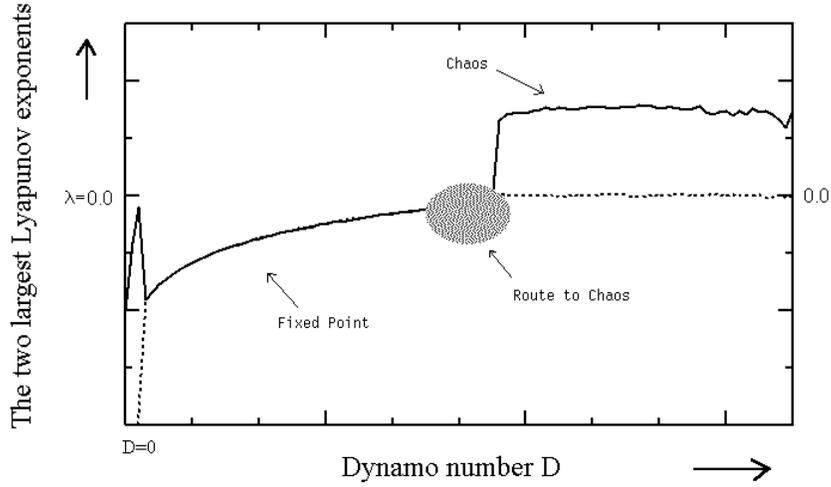}}
\caption[]{\label{chaos1}
Schematic graph of the typical asymptotic behaviour of the
two largest Lyapunov exponents for the $D>0$ case
with a driving term ${\cal F}({\bf B})\sim 
A_{\phi}B_{\phi}$. The route to chaos seems to vary 
as the truncation order increases.}
\end{figure}
%______________________________________________________________________

For the sake of comparison we summarise,
schematically, the results of the integration of the systems of 
Schmalz and Stix (see also Covas et al. 1996),
in Figure 1.
Here the largest Lyapunov exponent is depicted by a 
solid line and its negative, zero and positive values indicate 
equilibrium, periodic and chaotic regimes respectively (the second Lyapunov
exponent is plotted as a dashed line).

Since in many astrophysical settings
(including that of the Sun) the sign of the dynamo number, $D$,
is not known, we also study the effects of changing its sign.
The results for negative dynamo numbers
are given in Figure 2.

%______________________________________________________________________
\begin{figure}
%\vspace{7.2cm}
\centerline{\def\epsfsize#1#2{0.45#1}\epsffile[-50 60 800 580]{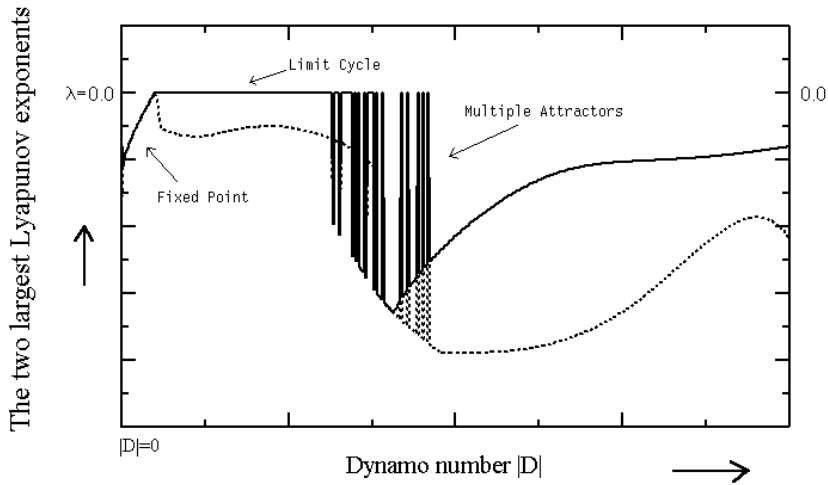}}
\caption[]{\label{chaos2}
Schematic graph of the typical asymptotic behaviour of the
two largest Lyapunov exponents for the $D<0$ case
with a driving term ${\cal F}({\bf B})\sim A_{\phi}B_{\phi}$.}
\end{figure}
%______________________________________________________________________

An interesting mode of behaviour occurs
in the spiky regions of Figure 2 which corresponds
to the presence of ``multiple attractors''
(of more than one attractor consisting of equilibrium and periodic
states) over substantial intervals of $D$.
 
To clarify the consequences of this behaviour, we have plotted in
Figure 3 the behaviour of the $N=4$ truncation as a function of small changes
in the dynamo number and the initial conditions. It clearly demonstrates that 
in these regimes small changes in either $D$ or the initial
conditions may produce drastic changes in the dynamical behaviour of the system.
This form of {\em fragility} could be of significance in producing seemingly
intermittent types of behaviour.
\\

%______________________________________________________________________
\begin{figure}
%\vspace{8cm}
\centerline{\def\epsfsize#1#2{0.45#1}\epsffile[0 0 612 512]{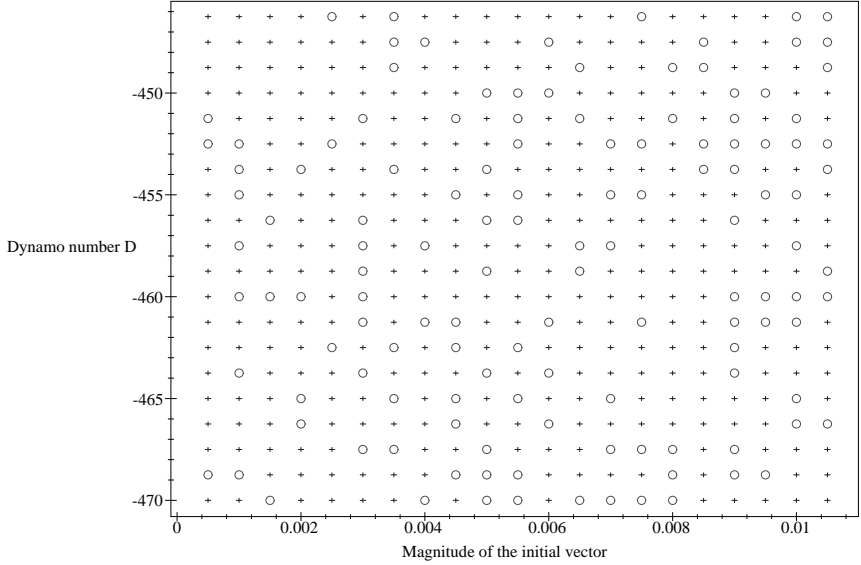}}
\caption[]{\label{fragility}
Fragility in the dynamics with respect 
to small changes in the dynamo number $D$ and
the magnitude of the initial vector $(A_{\phi}, B_{\phi}, \alpha_M)$.
Crosses and circles represent equilibrium and 
periodic behaviour respectively.}
\end{figure}

To study the robustness of this system we 
considered a modified feedback term in the form
\begin{equation}
{\cal F}({\bf B})=f_1 A_{\phi}B_{\phi} + f_2 \alpha|{\bf B}|^2,
\label{abalphab2}
\end{equation}
where $f_1=(\mu_0\rho)^{-1}$ and $f_2=(\mu_0\rho\beta)^{-1}$
($\beta$ is the combined (turbulent plus ohmic) diffusion of the
field, $\rho$ the density of the medium and $\mu_0$ the magnetic constant).
The additional second term on the right hand side
of (\ref{abalphab2}) is reminiscent of the type appearing in a more
physically motivated form of ${\cal F}$ given
by Kleeorin and Ruzmaikin (1982),
Zeldovich, Ruzmaikin and Sokoloff (1983)
and Kleeorin, Rogachevskii and Ruzmaikin (1995)
and discussed in Covas et al. (1996).

Figure 4 summarises the results for this modified form
of ${\cal F}$ and as can be seen, the inclusion of $\alpha|{\bf B}|^2$
in equation (\ref{abalphab2}) has drastic effects on the dynamics
of the system, for both positive and negative dynamo numbers.
In particular, it strongly suppresses the 
chaotic behaviour found by Schmalz and Stix.

%______________________________________________________________________
\begin{figure}
%\vspace{7.2cm}
\centerline{\def\epsfsize#1#2{0.45#1}\epsffile[-50 60 800 580]{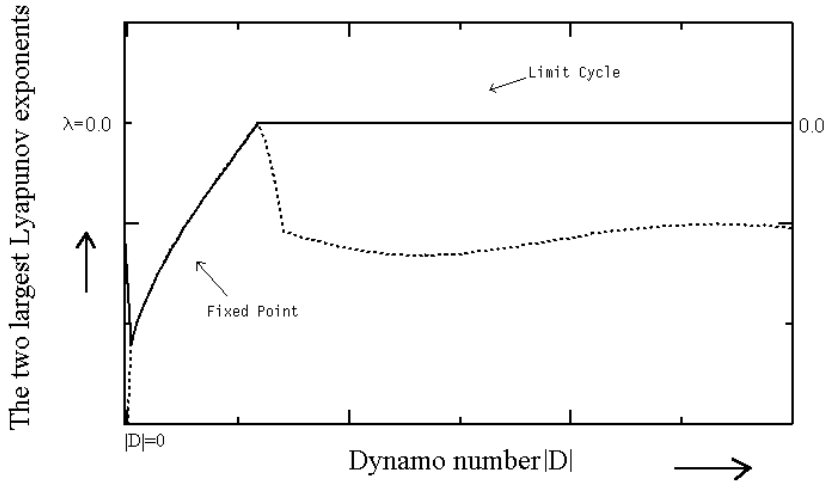}}
\caption[]{\label{cycle}
Schematic graph of the typical asymptotic behaviour of the
two largest Lyapunov exponents for the case with a driving term
${\cal F}({\bf B})=f_1 A_{\phi}B_{\phi} - f_2 \alpha|{\bf B}|^2$.
The overall behaviour is independent of the sign of D.}
\end{figure}
%______________________________________________________________________
\section{Comparison with other functional forms}
%______________________________________________________________________
In Covas et al. (1996), we considered the effects of using
more physically motivated choices for the driving term
given by  Kleeorin and Ruzmaikin (1982),
Zeldovich, Ruzmaikin and Sokoloff (1983) and Kleeorin, Rogachevskii and Ruzmaikin (1995)
in the form 
\begin{equation}
{\cal F}({\bf B})= g_1 {\bf J}\cdot{\bf B} + g_2 \alpha|{\bf B}|^2,
\end{equation}
where $g_1=-(\mu_0\rho)^{-1}$ and $g_2=(\mu_0\rho\beta)^{-1}$ are physical constants.

The results of considering the cases with
$g_1 \neq 0$, $g_2=0$ and $g_1 \neq 0$, $g_2 \neq 0$
can be summarised as follows. In the former case, the behaviour seems to
mirror that seen for the case with ${\cal F}({\bf B})\sim
A_{\phi}B_{\phi}$, in the sense that
for $D>0$ there is an asymptotic dominance
of ``multiple attractor'' regimes, while for $D<0$ this behaviour is
replaced by chaotic behaviour. In the latter case, however,
where the term proportional to $\alpha|{\bf B}|^2$ is included, there is a
suppression of both chaotic and ``multiple attractor'' regimes, and the
behaviour looks similar to that depicted in Figure 4.
We also changed the damping operator so that it was of the form
derived by Kleeorin, Rogachevskii and Ruzmaikin (1995).
It was found, however, that this change did not
produce any qualitative changes.

%______________________________________________________________________
\section{Conclusions}
%______________________________________________________________________
We have studied the robustness of truncated $\alpha \Omega$ dynamos 
including a dynamic $\alpha$ equation, with respect to 
a change in the driving term of the $\alpha|{\bf B}|^2$ type.
We find that this changes the results of Schmalz and Stix 
drastically by suppressing the possibility of chaotic behaviour.
Our results presented here and those in Covas et al. (1996)
show that changes in the driving term have
important effects on the dynamical behaviour of the resulting systems.
Furthermore, the sign of the dynamo number also plays
an important role, being capable of radically changing 
the behaviour of the system.
Typically we find that the behaviour of the system for
${\cal F}({\bf B}) \sim A_{\phi}B_{\phi}$ and $D>0$ is similar to the
one with ${\cal F} ({\bf B}) \sim {\bf J}\cdot{\bf B}$, but with $D<0$
and vice-versa.
The change of sign causes the
behaviour to change from typically chaotic to ``multiple attractor''
solutions. These correspond to regimes where equilibrium and
periodic regimes are present simultaneously.
As a result, small changes in $D$ or the initial conditions can
substantially change the behaviour of the system. This type of behaviour can
be of importance in producing intermittent type behaviour
observed in solar variability.

\acknowledgements EC is supported by grant  BD / 5708 / 95 -- 
Program PRAXIS XXI, from JNICT -- Portugal.
RT benefited from SERC UK Grant No. H09454. This research also benefited
from the EC Human Capital and Mobility (Networks) grant ``Late type stars:
activity, magnetism, turbulence'' No. ERBCHRXCT940483.
EC thanks the Astronomy Unit at QMW and the organisers of the
{\it 8th European Meeting on Solar Physics} for support to attend the
conference.

%______________________________________________________________________

%______________________________________________________________________

%______________________________________________________________________

\begin{thebibliography}{}
%______________________________________________________________________
\bibitem[1989]{betal}
Brandenburg, A., Krause, F., Meinel, R., Moss, D., Tuominen, I., 1989, A\&A, {\bf 213}, 411

\bibitem[1989]{btm}
Brandenburg, A., Moss, D., Tuominen, I., 1989, Geophys. Astrophys. Fluid Dyn., {\bf 40}, 129

\bibitem[1992]{coley}
Coley, A.A.,  Tavakol, R.K., 1992, Gen. Rel. Grav., {\bf 24}, 835

\bibitem[1996]{us}
Covas, E., Tworkowski, A., Brandenburg, A., Tavakol, R., 1996, A\&A,
(to be published)
 
\bibitem[1992]{feudel92}
Feudel, F., Jansen, W., Kurths, J., 1993, Int. J. Bifurc. Chaos, {\bf 3}, 131

\bibitem[1983]{gilman83}
Gilman, P.A., 1983, ApJS, {\bf 53}, 243

\bibitem[1982]{klrz}
Kleeorin, N. I., Ruzmaikin, A.A, 1982, Magnetohydrodynamica,{\bf  N2}, 17

\bibitem[1995]{krr}
Kleeorin, N. I, Rogachevskii, I., Ruzmaikin, A., 1995, A\&A, {\bf 297}, 159

\bibitem[1980]{krause80}
Krause, F., R\"adler, K.-H., 1980, {\it Mean-Field Magnetohydrodynamics and
Dynamo Theory}, Pergamon, Oxford

\bibitem[1991]{SchmalzStix91}
Schmalz, S., Stix, M., 1991, A\&A, {\bf 245}, 654

\bibitem[]{smale}
Smale, S. 1966, Amer. J. Math., {\bf 88}, 491

\bibitem[1988]{ellis}
Tavakol, R.K., Ellis, G.F.R.,1988, Phys. Lett., {\bf 130A}, 217
 
\bibitem[1995]{tt}
Tavakol, R.K, Tworkowski, A. S., Brandenburg, A.,
Moss, D., Tuominen, I., 1995, A\&A, {\bf 296}, 269

\bibitem[1995]{twk95}
Tobias, S.M., Weiss, N.O., Kirk, V., 1995, MNRAS, {\bf 273}, 1150

\bibitem[1984]{cwj84}
Weiss, N. O., Cattaneo, F., Jones, C. A., 1984, Geophys. Astrophys. Fluid Dyn., {\bf  30}, 305

\bibitem[1993]{zeldovich83}
Zeldovich, Ya.B., Ruzmaikin, A.A, Sokoloff, D.D,
1983, {\it Magnetic Fields in Astrophysics}, Gordon and Breach, New York

%______________________________________________________________________
\end{thebibliography}
\end{document}